\documentclass{pas}

\usepackage{multirow}

\usepackage{comment}

\newcommand{\tblu}{\color{black}} 
\newcommand{\tbla}{\color{black}}

\begin{document}

\lefttitle{Publications of the Astronomical Society of Australia}
\righttitle{Cambridge Author}

\jnlPage{1}{4}
\jnlDoiYr{2021}
\doival{10.1017/pasa.xxxx.xx (pre-submission)}

\articletitt{Research Paper }











\title{GBD-DART-II: 175 MHz Polarimetric Observation of Pulsars from Gauribidanur and a New Pulsar Signal Processing Pipeline }

\author{\gn{Arul Pandian B}$^{1,2}$, \gn{Joydeep Bagchi}$^{1}$, \gn{Prabu Thiagaraj}$^{2}$, \gn{K.B.Raghavendra Rao}$^{2}$, \gn{Vinutha Chandrashekar}$^{2}$}

\affil{$^1$The Christ (Deemed to be University), Bangalore., $^2$Raman Research Institute, Bengaluru, India. }

\corresp{Arul Pandian B, Email: arulpandian022@gmail.com}

\citeauth{Author1 C and Author2 C, an open-source python tool for simulations of source recovery and completeness in galaxy surveys. {\it Publications of the Astronomical Society of Australia} {\bf 00}, 1--12. https://doi.org/10.1017/pasa.xxxx.xx}

\history{(Received xx xx xxxx; revised xx xx xxxx; accepted xx xx xxxx)}

\begin{abstract}
 
A new pulsar signal-processing pipeline has been developed for observing pulsars with the Diamond Array Radio Telescope (DART) at the Gauribidanur radio observatory (13.604 N, 77.427 E). The array consists of 32 off-axis dual-polarised LPDAs, with a nominal gain of 22 dBi between 130 and 350 MHz and a 15-degree HPBW at 175 MHz, and it performs transit observations on pulsars. Custom-developed analogue signal conditioning, transmission, and a digital backend are incorporated to record the data. A real-time data-capture and analysis tool has been developed that receives band-limited time-series data and outputs voltages from a transient buffer, as well as full-polar spectral data at both high and low resolutions, suitable for transient searches and pulsar studies. Additionally, full-polar folded profile archives are generated for known pulsars in subintegrations and both coherent and incoherent dedispersion. An all-day monitoring is implemented that continuously records a one-second averaged spectrum. Custom-developed Python routines, FFT libraries, DSPSR, PSRCHIVE, and Presto modules have been used to build the pipeline. The functionalities of the pipeline were validated with artificially generated pulsar signals and strong celestial sources before it was released for routine observations. Presently, the pipeline is configured to observe pulsars between 170 and 196 MHz, with a daily cadence. Recorded data are reduced in-line immediately following each observation, nearly matching the observation time at a 1:1 ratio. An Intel i9 server captures the data, and an AMD R9 CPU does the primary data reduction. The archives are routinely backed up to a remote system via the internet. The paper presents the architecture of the signal processing pipeline developed, its validation, and initial polarimetric results observing five bright pulsars at 175 MHz. Results also include RM estimates and single-pulse study results for B0953+08, B0531+21, and B1133+16, as well as from monitoring the spin-down of the Crab pulsar over 200 days of observation. Finally, it presents a discussion on the potential improvements for the array.

\end{abstract}


\maketitle

\section{Introduction}
 
 Low-frequency radio pulsar observations are crucial for understanding pulsar emission mechanisms, galactic magnetic fields from the observed Faraday rotation \citep{sobey2017studying}, and the interstellar medium via scattering and dispersion measurements. The recent efforts to detect low-frequency gravitational waves utilise pulsar timing arrays (PTAs) \citep{fos90}, formed by a collection of millisecond pulsars (MSBs) in our galaxy. Continued observations and correlation of the pulse time-of-arrival (ToA) errors from the MSBs located at vantage positions serve to implement a galactic-scale interferometer and aid in investigating disturbances to the space-time metric due to a passing gravitational wave \citep{edw06}. Precise knowledge of the dispersion measure (DM) is essential for accurate ToA estimation of a pulsar. The DM estimation from low-frequency observations yields accuracies an order of magnitude greater than with higher-frequency observations alone \citep{kri21}. Such observations also provide our most direct clues to the structure of the magnetic fields around pulsars. They are essential in the case of millisecond pulsars, where radio emission necessarily originates very close to the surface of the neutron star. 

The linearly polarised emission in the rotating vector model (RVM) tracks changes in the dipolar magnetic field line planes, which causes the polarisation position angle (PA) swing \citep{radhakrishnan1969magnetic}. Simulations \citep{foster2015intrinsic} show that the polarisation characteristics of the pulsar can be utilised to enhance the ToA estimation. An estimate of a telescope's instrumental responses can be obtained from full-polarisation observations of known bright pulsars. By employing polarimetric data, the ToA estimates can be improved \citep{susarla2025long, guillemot2023improving}. 

The Gauribidanur radio observatory facilitated single-polarisation pulsar observations across a 1.5 MHz band at a centre frequency of 34.5 MHz \citep{deshpande1992pulsar} and also between 60 and 80 MHz \citep{bane2024initial}. The current work enables instantaneous 16 MHz-wide dual polarised pulsar observations between 130 and 350 MHz using the GBD-DART pulsar array \tblu[Pandian,2025 in review].

The design details of the GBD-DART and its capabilities to observe solar transients and pulsars are given in \tblu[Pandian,2025 in review]\tbla.  In this paper, we provide details of the newly developed pulsar observing capabilities and the results from observing a set of bright pulsars in a full polar mode. 

The LOFAR, LWA, and MWA telescopes use specific pulsar processing pipelines based on PSRCHIVE and Presto utilities \citep{stappers2011observing, bondonneau2021pulsars, stovall2015pulsar, bhat2023southern, yu2025study, ran11}. The current work uses a combination of custom-developed Python scripts for data capture and PSRCHIVE and Presto modules for near-real-time data reduction and archive generation.

\section{Pulsar Signal Processing}

We present a brief overview of the array architecture and the signal conditioning in this section. 

\subsection{ Overview }
 
\begin{figure}[ht]
    \centering
    \includegraphics[width=0.7\linewidth]{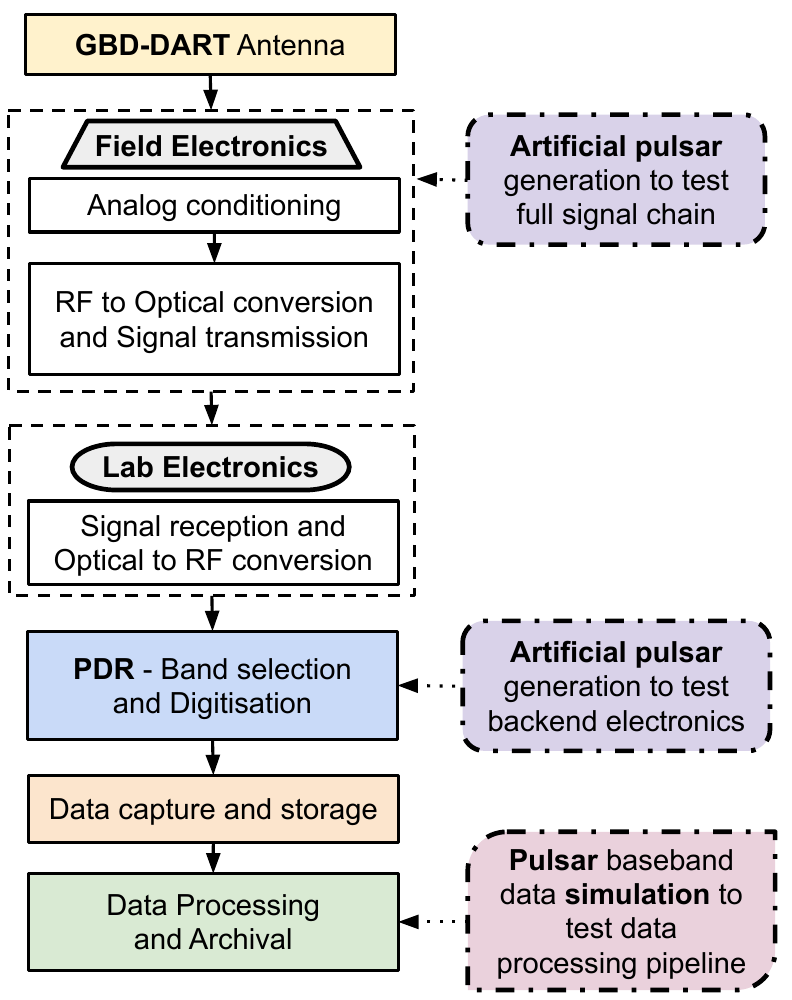}
    \caption{GBD-DART Signal Chain, data flow, and testing provisions of the pipeline.}
    \label{rf_path}
\end{figure}

An overview of the GBD-DART antenna array signal chain is given in Figure \ref{rf_path}. The array as outlined in Section \ref{sectANT} consists of 32 off-axis dual-polarised LPDAs configured as a diamond-shaped tile. The RF signals from each polarisation are suitably amplified and combined independently to create a voltage beam towards the zenith. Then the combined signals are further amplified and transmitted as optical signals to the observatory lab, located 300 metres away, where they are converted back to RF signals. Details of the analogue front-end system are presented in Section \ref{sectADB}. The RF signals are then band-limited to 16 MHz using a set of band-pass filters, translated to an IF band in a heterodyne receiver, and band-pass sampled with a 33 MHz sampling clock. Further details of this system are presented in Section \ref{sectPDR}.

 Sampled data is packetised and sent through Ethernet to a data recording system. The data recording system incorporates transient buffers, data recording, and health monitoring features. During a pulsar observation, the transient buffer data is passed to a data processing system, where the recorded data is processed immediately after each observation using a combination of custom and standard pulsar data reduction utilities. The reduced archives after processing are subsequently backed up to a remote storage system, while the raw data (approximately 220 GB per hour of observation) is not retained. Further details of this data processing are presented in Sections \ref{sectPDF} and \ref{sectPDPP}.

The receiver chain also incorporates tools to generate artificial pulse signals, aiding in the diagnosis of critical pipeline stages. The use of artificial signal generation is described later in section \ref{sectSV}. 

\subsection{Antenna Array} \label{sectANT}
GBD-DART is a diamond-shaped array formed out of 64 log-periodic dipole arrays (LPDA) positioned in a checker-board layout. The LPDAs are grouped in four, with opposite pairs slightly tilted toward each other, thus forming a pyramid shape. This primary configuration provides dual-polarised outputs with a half-power beam-width (HPBW) of between 60 and 70 degrees for the 130 and 350 MHz bands when the signals from the opposite pairs of LPDAs are combined. The pyramids have a maximum baseline of 7 meters in the array. The signals from the pyramids are combined in a tree fashion to give the array gain of approximately 22 dBi at 200 MHz (about 15 degrees HPBW).

\subsection{Analog Front-end}\label{sectADB}
 A current BALUN arrangement feeds each LPDA output through a low-noise amplifier (LNA).  The LNAs provide 20 dB of gain and are fitted with a high-pass filter at the inputs to reject out-of-band signals below 130 MHz, and a Bias-T network at the LNA outputs to feed a DC supply to the LNAs.

\begin{figure}[ht]
    \centering
    \includegraphics[width=\linewidth]{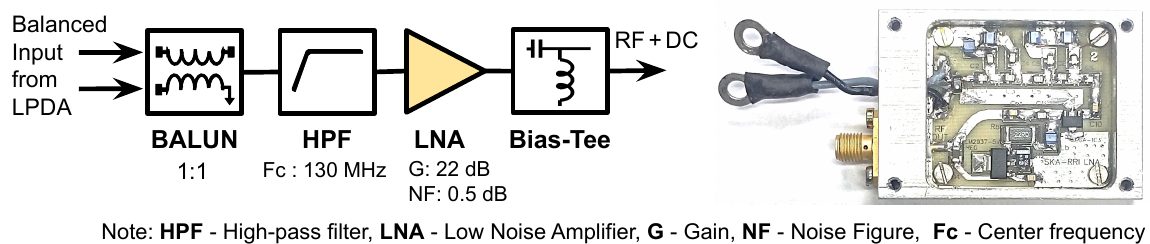}
    \caption{Differential signals from the LPDA are fed to the Low Noise Amplifier via a BALUN. The amplifier inputs are fitted with a high-pass filter to reject frequencies below 130 MHz. The low noise amplifier module has a gain of 20 dB and a noise figure of 1.35 dB.}
    \label{Tubular_Rx}
\end{figure}

Since the LNA outputs are to be combined to form a phased array output, the individual LNA gains are matched, the phases of the LNAs, interconnecting cables, and combiners are aligned to be within about $\pm5^{\circ}$. The voltage sum from the two polarisations (32 LPDAs each) is routed to a high-fidelity analog signal conditioning front-end module, the Tubular Receiver.  

The Tubular Receiver, as outlined in Figure.\ref{Tubular_Rx}, consists of three filters, three amplifiers and four attenuators that are interleaved suitably to yield a gain of about 51$\pm{0.5}$ dB between the 120 to 350 MHz band.

\begin{figure}[ht]
    \centering
    \includegraphics[width=\linewidth]{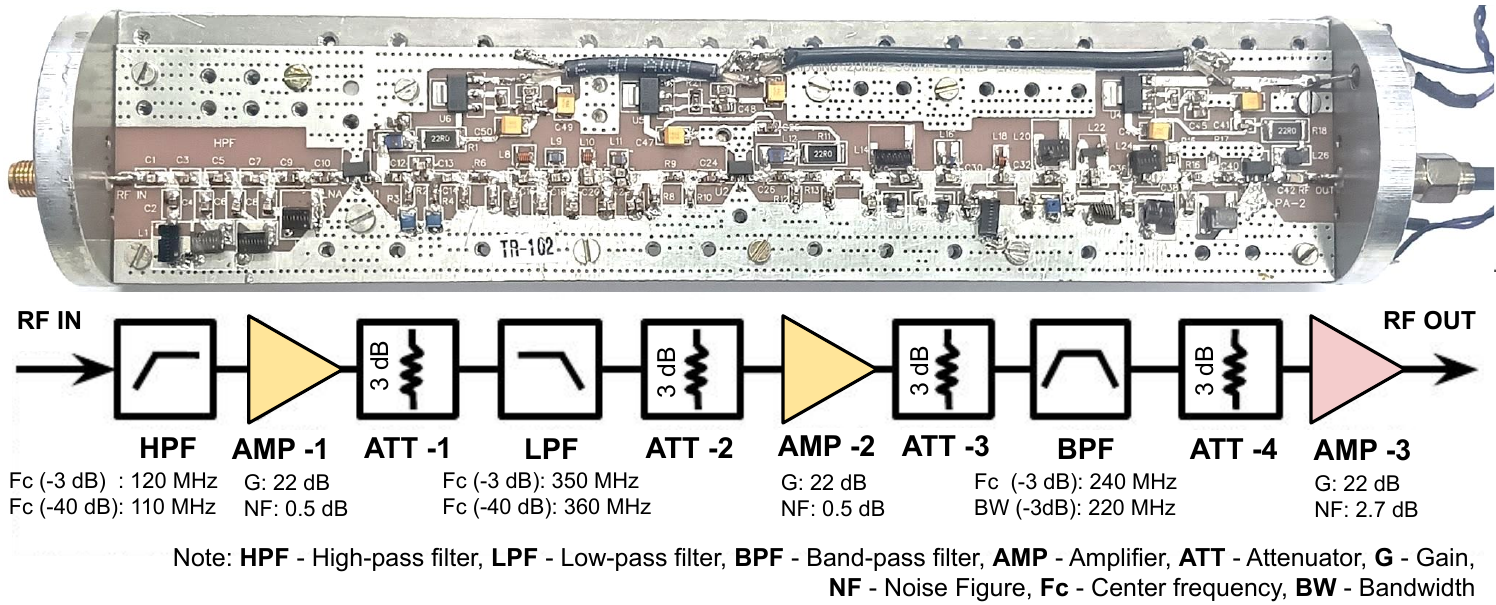}
    \caption{Picture show the analog conditioning module, the Tubular Receiver circuit board and it's functional blocks. It consists of three stages of filtering (HPF, LPF and BPF) and three stages of amplifiers (AMP-1 to AMP-3). The filters are preceded and followed by a 3 dB attenuators to maintain operational stability.}
    \label{Tubular_Rx}
    
\end{figure}

 The amplified signals from the Tubular Receiver are routed to an electrical-to-optical converter module, which transmits the two polarisation radio bands as an optical amplitude-modulated signal at 1310 nm through a single-mode fibre to the observatory electronics lab, a distance of 350 m.   The optical signals are converted to electrical signals at the observatory electronics room and fed to a portable digital receiver back-end for digitisation and recording.


\subsection{Analog and Digital Back-end} \label{sectPDR}
\begin{figure}[ht]
    \centering
    \includegraphics[width=0.99\linewidth]{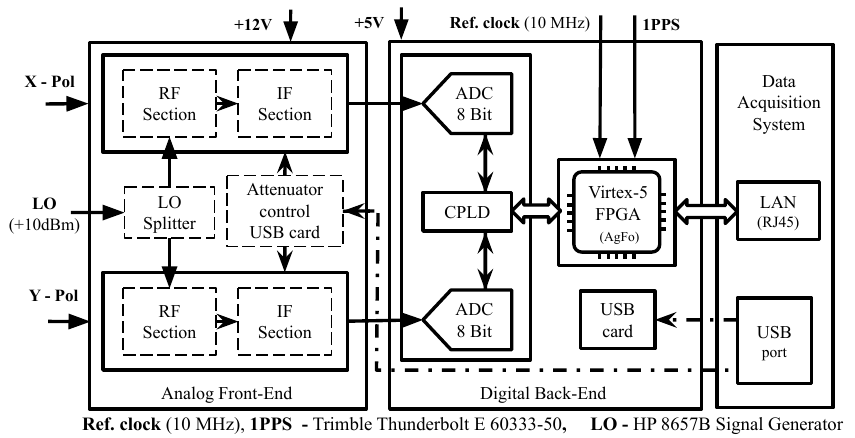}
    \caption{ The portable dual receiver (PDR) is a two-channel, 16-MHz-wideband heterodyne receiver and digitiser.}
    \label{PDR_block_diagram}
\end{figure}

The analog and digital back-end, as outlined in Figure \ref{PDR_block_diagram}, is  a portable dual receiver (PDR) with two-channel 16-MHz wide-band heterodyne receiver cum digitiser \citep{vinutha2017miniaturised}. It can select a contiguous 16 MHz band from the 130 and 350 MHz span, using a suitable local oscillator frequency. The local oscillator clock for the mixer stage of the receiver is generated using an HP synthesizer. An outline of the RF and the IF internal sections of PDR is given in appendix, Figure \ref{PDR_RF_IF}. The sampling clock for the digitiser is generated using a DLL in the FPGA. Both the HP synthesiser and the FPGA receive a 10 MHz reference clock generated from a Trimble make GPS unit. A 1-PPS signal from the Trimble synchronises start of data flow from the receiver. The PDR has two 8-bit digitisers sampling the two  polarisation signals at 33 MSPs. The sampled data from the two digitisers are time-tagged, packaged into a UDP packet, and transmitted via Ethernet ports to a data acquisition system (DAS). The PDR uses analog device ADCs for digitisation, and an Xilinx Virtex-5 FPGA for packetising the sampled data and transmitting time-tagged data packets over a Gigabit Ethernet port. 

\subsection{Data Capture} \label{sectPDF}
The UDP packets arriving from the PDR are captured in the ramdisk by the data acquisition system (DAS) using the GULP utility \citep{coreySatten2008}. The GULP writes out the UDP packets in a PCAP format \citep{harris2025pacp}. The DAS maintains three ring buffers of sizes 10, 20, and 70 GB to facilitate the smooth transfer of observation data from Ethernet to the spectrum monitoring and transient data recording processes. A set of dedicated buffer management routines orchestrates the data copying tasks across the buffers. The 10 GB ring buffer (the GULP buffer) eases the arrival of UDP packets by accumulating them in 2 GB-sized files, once every 30 seconds, corresponding to a 2x33 MSPS data rate.

\begin{figure}[ht]
    \centering    \includegraphics[width=0.75\linewidth]{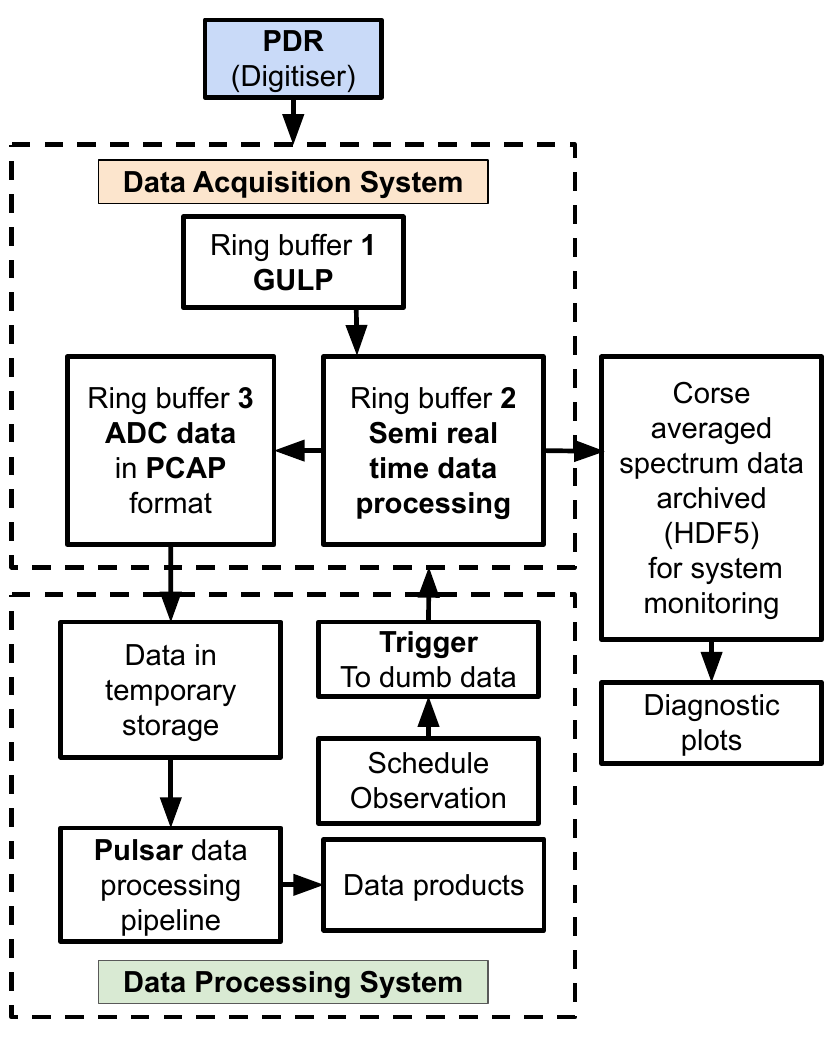}
    \caption{Data Capture Architecture}
    \label{data_capture_architecture}
\end{figure}

The 2 GB files from the GULP  buffer are swiftly transferred to the 20 GB  staging buffer, which is then read by a processing task to compute the two self, cross and phased sum products for the two ADC input channels. These products are archived in HDF5 format in the DAS server with a spectral resolution of 64.5 kHz and a time resolution of 1 s, and are continuously recorded 24/7 to monitor the system's performance. The data from the 20 GB ring buffer is also copied to the 70 GB buffer, also known as a transient buffer. The role of the transient buffer is to always maintain the last 5 (or programmable number of)  minutes of raw data (10x2 GB) files so that an observation for a pulsar or a transient source such as an FRB could be carried out instantly when a trigger is received from an external system. During a pulsar observation, a trigger is sent to thr DAS. Upon receiving the trigger command, the DAS PC would send a programmed number of  2  GB files ( 120 for an hour-long observation) from the transient buffer to an external PC through a Gigabit Ethernet port. The DAS PC connects to the PDR through a dedicated 1-Gigabit Ethernet port to facilitate streaming data flow. An outline of this data capture architecture is outlined in figure \ref{data_capture_architecture}.

\subsection{Pulsar Data Processing} \label{sectPDPP}
\begin{figure}[ht]
    \centering
    \includegraphics[width=0.75\linewidth]{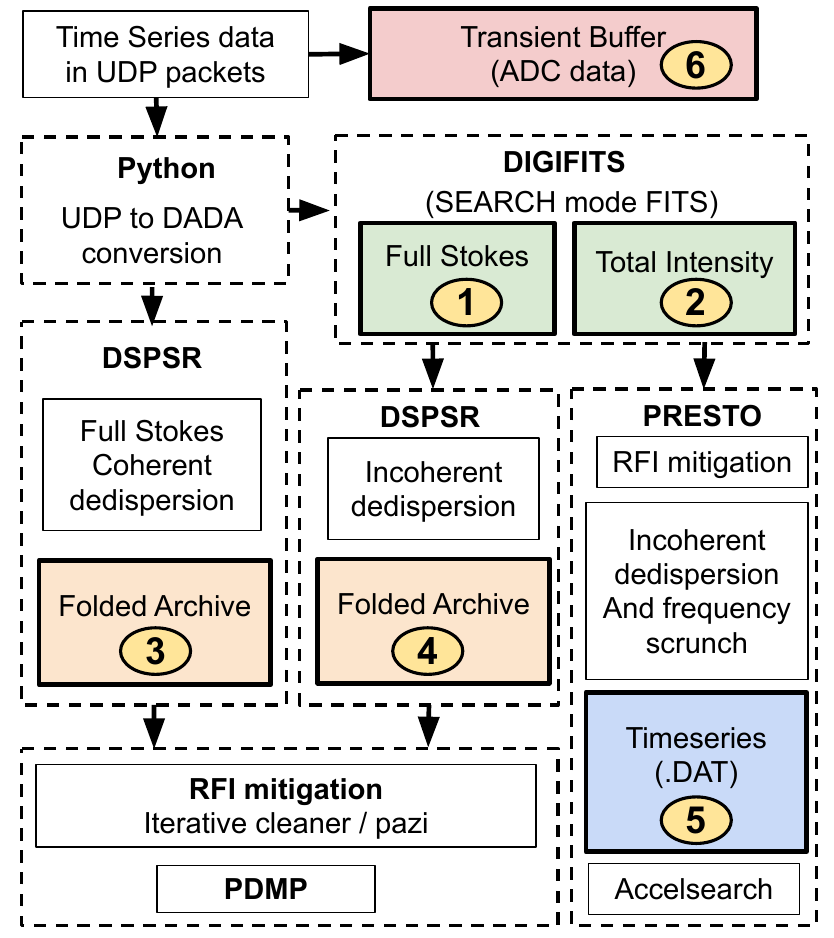}
    \caption{Pulsar Data processing pipeline}
    \label{pipeline}
\end{figure}

The digitised observation data acquired with GULP is in a PCAP format. A Python script was used to count the packets and remove the PCAP header from the UDP packets. Each data packet will contain 512 samples, alternating between the two channels. Time series data were written in the DADA \citep{van21} format, with the first 4096 bytes containing the observation settings. Depending on the choice of data processing, single-polarisation or full-polarisation data can be written out. Multiprocessing \citep{mck12} libraries from Python were used to accelerate the DADA conversion. As shown in Figure \ref{pipeline}, data was processed in coherent de-dispersion with DSPSR \citep{van11} and  DIGIFITS, which is part of DSPSR, to generate total-intensity and full stokes search mode PSRFITS. further data can be used to perform Incoherent de-dispersion and single pulse search. Coherent de-dispersion mode, dspsr performs a long FFT depending on the given dispersion measure (DM) of the pulsar and deconvolves the response of the interstellar medium, then performs IFFT. This time series data was again Fourier-transformed using a given number of FFT points, and the power spectra were averaged and folded with pulsar periods from the given ephemeris file taken from the ATNF pulsar catalogue \citep{hob04}.

DSPSR performs Incoherent de-dispersion and folding on the filter-bank data generated by \textit{digifil}. Another copy of the filter-bank data is processed by PRESTO \citep{ran11} utilities. The \textit{rfifind} was used to perform RFI mitigation, \textit{prepdata} and \textit{accelsearch} were used to de-disperse, frequency scrunch and search for pulsar candidate in the data. \textit{Prepfold} utility was used to fold the pulsar candidate and predict the period and period-derivative of the pulsar. PSRCHIVE utilities \citep{hot04} were used to inspect the Folded archive in FITS format. The iterative cleaner and \textit{pazi} were used to perform RFI cleaning in fits files. The cleaned archives are used to search in period, period-derivative and DM space to estimate best values using \textit{pdmp}.  

\section{ System Validation} \label{sectSV}

The new pipeline was validated with artificial pulsar signals. A test setup, as outlined in Section \ref{sectEPSR}, was used to study the instrument's dynamic range and to determine a linear operating range suitable for pulsar observations. Similarly, fake pulsar polarimetric data, as described in Section \ref{sectSPSR}, was generated and used to validate the full polar pulsar data reduction software pipeline developed.  

\subsection{Validation with Emulated Pulse Signals}\label{sectEPSR}

\begin{figure}[ht]
    \centering 
    \includegraphics[width=0.9\linewidth]{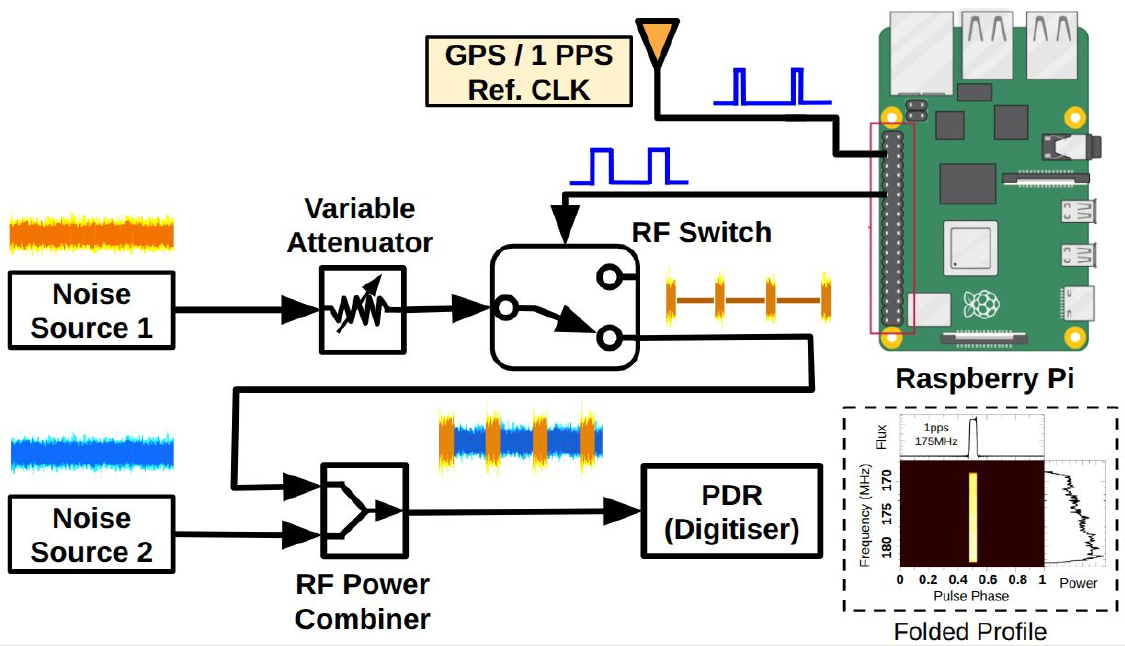}
    \caption{Pulsar emulation signal flow. }
    \label{Artti_psr}
\end{figure}

A pulsed signal generator \citep{abhishek2025artificial}, as shown in Figure \ref{Artti_psr}, built based on two noise sources, an RF switch, a power combiner, and a \textit{Raspberry Pi}, was used to generate artificial pulsar signals. The Raspberry Pi module's role is to receive a one-pulse-per-second (1PPS) signal from a GPS and initiate the pulse generation by controlling the duty cycle of the RF switch. The instrument setup allows on- and off-pulse power levels to be varied over a broad range to study the recovered SNR from the system. 

The two noise source signals combine during the on-pulse period (when the RF switch is ON), determining the pulse width and pulse power. The second noise source (2) signal, which is always present, determines the off-pulse power. During the study, the off-pulse power was manually set to -70 dBm, and the on-pulse power was varied around -70 dBm using the variable attenuator in 3 dB increments over a 50 dB range, while maintaining a constant pulse width of 100 ms and a pulse repetition period of 1 s.

\begin{figure}[ht]
    \centering
    \includegraphics[width=\linewidth]{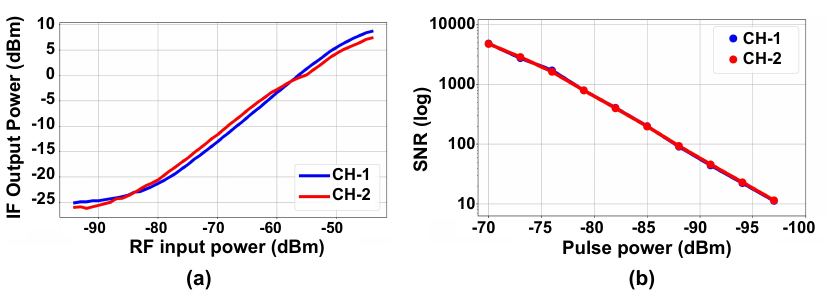}
    \caption{(a) PDR linearity test to find optimum operating RF power. (b) SNR linearity test on PDR channals with emulated pulsar signal shown in Figure \ref{Artti_psr}.  }
    \label{SNR_test}
\end{figure}

 The artificial pulse signals for the different pulse-power settings were passed through the analogue chain and then into the PDR for digitisation and recording. The data collected over 100 seconds for each setting were Fourier-transformed and then averaged by folding at the 1s pulse period across a fixed number of bins using standard pulsar processing tools. Figure \ref{SNR_test}a shows the relationship between the input noise power (RF input power: sum of two noise source signals) fed to the analog chain and the power at the IF output fed to the digitiser, reflecting the observed linearity. The SNR observed in the folded pulses for each of the input pulse power settings is shown in Figure \ref{SNR_test}b,  reflecting the consistency after digitisation. 

This test was also repeated on each RF path, including the RF optical modules, to verify their consistency and efficiency in processing pulsed signals. 

\subsection{Validation with Simulated Pulse Polarisation }\label{sectSPSR}

We studied the consistency of the software pipeline for measuring polarised signals using simulated signals. We used the Stokes pulse profile from the EPN catalogue as a template \citep{gould1998multifrequency} to obtain the constituent amplitudes and phase differences required for the X- and Y-polarisation voltage signals of the measured coherence. 

\begin{figure}[ht]
    \centering
    \includegraphics[width=0.7\linewidth]{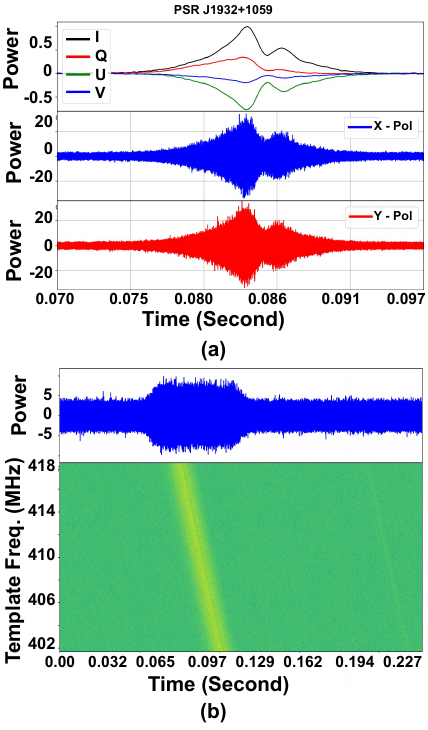}
    \caption{Polarised pulse simulation, (a) On pulse full stokes PSRFITS template profile of J1932+1059 pulsar at 410 MHz (top panal), Computed time series of the X - polarisation (middle panal), Computed time series of the Y - polarisation (bottom panal). (b) Dispersed time series of the X -polarisation. }
    \label{pulse_simulation}
\end{figure}

\begin{figure}[ht]
    \centering
    \includegraphics[width=0.8\linewidth]{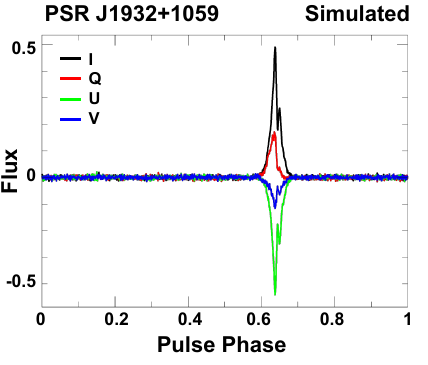}
    \caption{Pulse profile recovered by the pipeline after processing the simulated EPN B1929+10 profile data shown in figure. \ref{pulse_simulation}.}
    \label{pulse_after_proc_simulated_data}
\end{figure}

  Random integers were generated as 8-bit ADC samples with 4 sigma level using Numpy \citep{har20}.  These samples are created with a rate of 33 MSPs, the same as the digitiser's sampling rate.  The pulsar profile with complete Stokes information was used from the EPN pulsar catalog.  A 1024-bin profile from the EPN catalogue was interpolated to yield 33 noise samples.
 
To create timeseries voltage samples, this 2-D template was utilised in conjunction with the \textit{Noise generator} package from \textit{baseband-tasks} \citep{mar25b}.  These samples are later dispersed using the \textit{Disperse} package from \textit{baseband-tasks}.  \textit{Numpy} \citep{har20} was used to scale dispersed timeseries data to integers, produce 8-bit ADC samples with a 4 sigma level, and write out in a \textit{DADA} format \textit{baseband} \citep{mar25a}.

This voltage time series was dispersed for our study by using the catalogued  410 MHz DM, shown in Figure \ref{pulse_simulation}b. In Figure \ref{pulse_simulation}a, the Template full-Stokes pulsar profile is shown in the top panel. The middle and bottom panels show the X- and Y-polarisation voltage time series generated from the coherence product of the template data. 

The simulated data were analysed using the processing pipeline to generate the polarisation end products shown in Figure \ref{pulse_after_proc_simulated_data} that match well to the EPN profile.\tbla

\subsection{Validation of Signal Integrity}

The signal processing pipeline of the DART system (B Pandian, 2025) hardware modules were verified for signal integrity before commencing the regular pulsar observations. We highlight here several key factors that were taken into account.
 
The DART antenna elements are designed in an off-axis configuration with a 23-degree zenith tilt, resulting in approximately twice the gain of a single LPDA with the same physical area. The low-noise amplifier (LNA) and the balun are mounted at the antenna feed point to avoid resistive losses, which are crucial to determining the receiver chain's noise floor. Individual LNAs are tuned and phase-matched to the reference LNA to minimise signal loss during phased addition. We have used LMR-195 RF cable, which is certified for outdoor use and UV-resistant. Each cable was precisely cut, manually crimped, and tested for attenuation and phase match. The 8-way and 4-way power combiner was designed with ADP-2-1W RF power combiner chips and phase-matched to combine all 64 LPDAs of the array in two stages. From LNA to the power combiner, we have maintained the phase coherence with the tolerance of $\mp$5 degrees.

In the field, analog conditioning was performed using an analog tubular Rx, which consists of high-pass and low-pass filters and three-stage high-gain amplifiers, as shown in Figure \ref{Tubular_Rx}. The amplified RF signal was sent through the RFoF modules, which are enclosed in aluminium boxes that provide RF shielding. An RF signal leakage test was conducted between the X-pol and Y-pol channels using an artificial pulse generator setup to verify that better than -20dB isolation is achieved. The dynamic range of the analog and digital receivers was estimated using a linearity test. To avoid unwanted pickup and improve the isolation of the digitiser, we have replaced RG174 with higher-shield semi-flex cables in the PDR analog sections. The power supply of the PDR digitiser was changed from an SMPS to a linear supply to avoid switching noise on the power rails. A GPS-disciplined 1PPS and 10 MHz reference clock generator (Trimble Thunderbolt E 60333-50) was used to lock the HP 8657B Signal Generator, which served as a local oscillator (LO). The HP 8657B Signal Generator was tested with a frequency counter to verify the accuracy of the clock.

To ensure we do not lose the data during streaming UDP Ethernet transfers, we have used a RAM disk as a ring buffer to handle the incoming data rate (66MSPs). A three-stage RAM-disk ring buffer was used to process array monitoring data in parallel. 

\section{Results and Discussions}

This section presents results from observing five pulsars. The data presented here were obtained in full Stokes mode of observation, using the IAU convention for polarimetric observations. Polarisation calibration of the data was performed for the receiver response using the \textit{pac} package from PSRCHIVE. An instrumental calibration solution was obtained by feeding an artificial pulsed noise signal at the first-stage analog receiver in the field.

\subsection{J0953+0755}
\begin{figure}[ht]
    \centering
    \includegraphics[width=\linewidth]{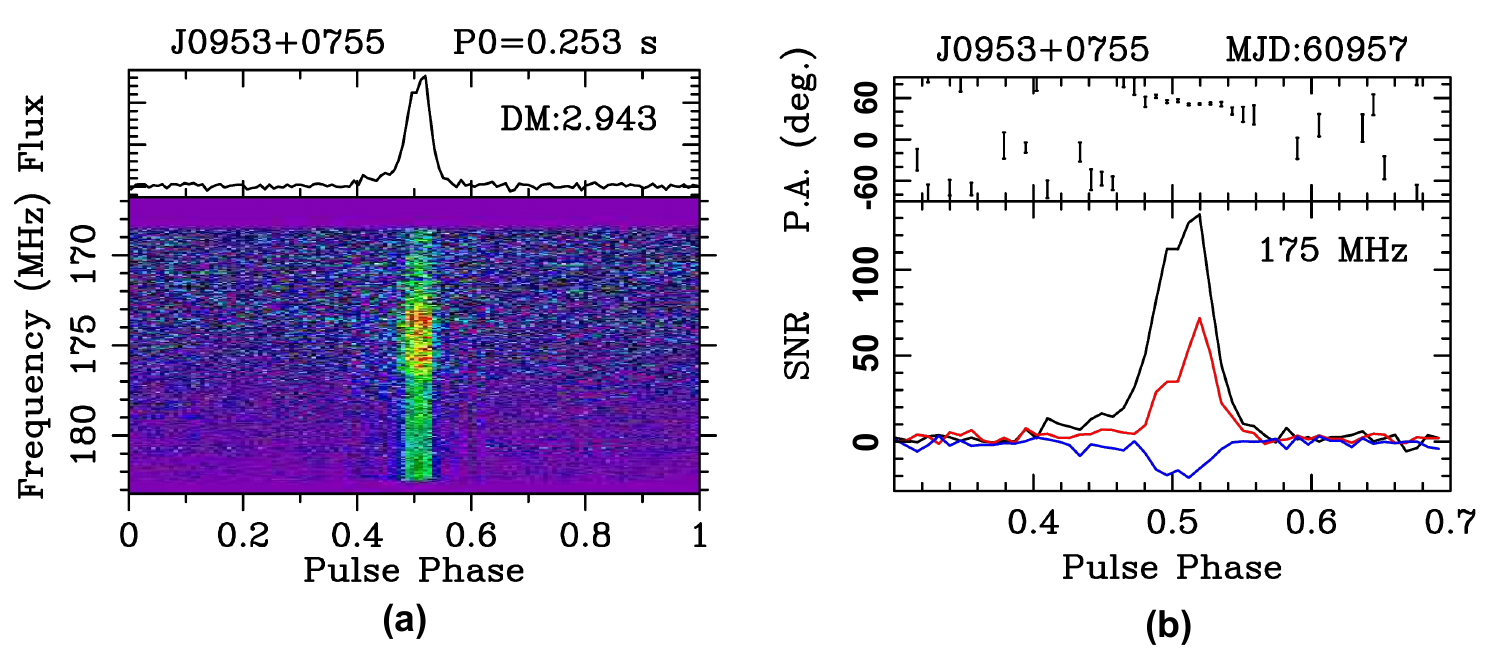}
    \caption{Pulse profile of J0953+0755. (a) Pulse intensity across the frequency. (b) Polarisation fraction, total intensity (Black), linear polarisation (Red), circular polarisation (Blue).} 
    \label{J0953+0755_profile}
\end{figure}

PSR J0953+0755 (B0950+08) appears to be a solitary pulsar with a period of 253.07 milliseconds and a low dispersion measure of 2.969 $pc/cm^3$. This pulsar is one of the brightest and most highly polarised, and it is routinely studied for giant pulses, single-pulse variability, and its intrinsic polarisation properties to better understand the radio emission region and provide constraints on the pulsar model \citep{tan2025detailed}. In Figure \ref{J0953+0755_profile}a, we present a high SNR average pulse (AP) profile and pulse intensity across the frequency from one of our 40 minutes of observation. The fraction of polarisation with position angle is shown in Figure \ref{J0953+0755_profile}b.

\begin{figure}[ht]
    \centering
    \includegraphics[width=\linewidth]{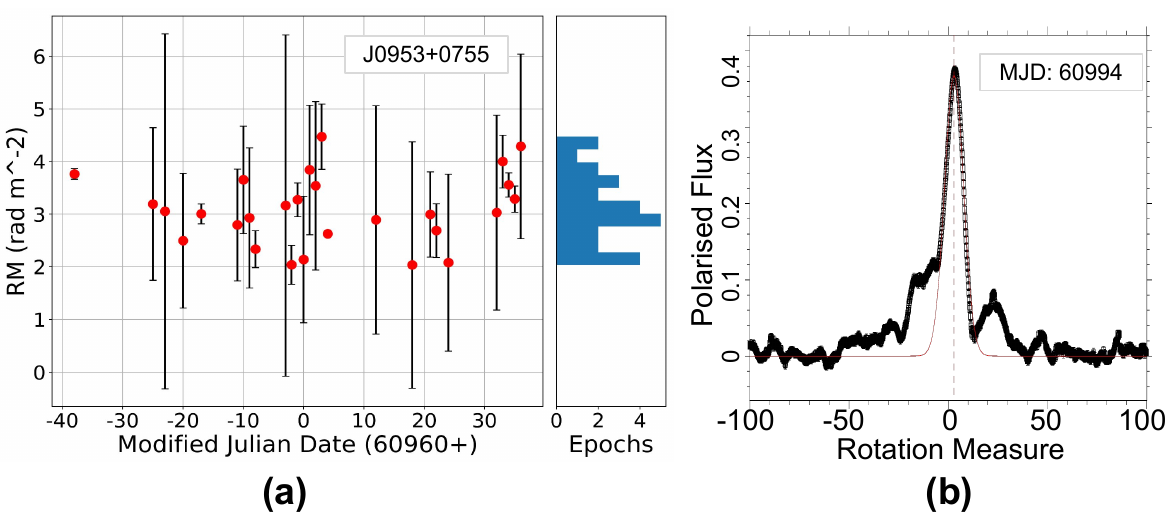}
    \caption{J0953+0755 Rotation measure estimation using rmfit. RM = 3.08$\mp$ 1.14 $rad/m^2$} 
    \label{J0953+0755_RM_plot}
\end{figure}

In low-frequency radio observations, this pulsar is being studied extensively to investigate the effect of interstellar scintillation on pulse profile variability over frequency \citep{shabanova2004properties,ziwei2022pulsar}. In Figure \ref{J0953+0755_RM_plot}a, we have presented the estimated rotation measure (RM) in a span of 100 days from our data. The \textit{rmfit} package from PSRCHIVE was used to calculate the RM. Figure \ref{J0953+0755_RM_plot}b illustrates RM measurement being 3.08$\mp$1.14 $rad/m^2$ for MJD 60994.

The pulse amplitude of this pulsar was highly variable and known for generating high-intensity giant pulses with a factor of more than 200 \citep{smirnova2012giant}. 
 
\begin{figure}[!ht]
    \centering
    \includegraphics[width=\linewidth]{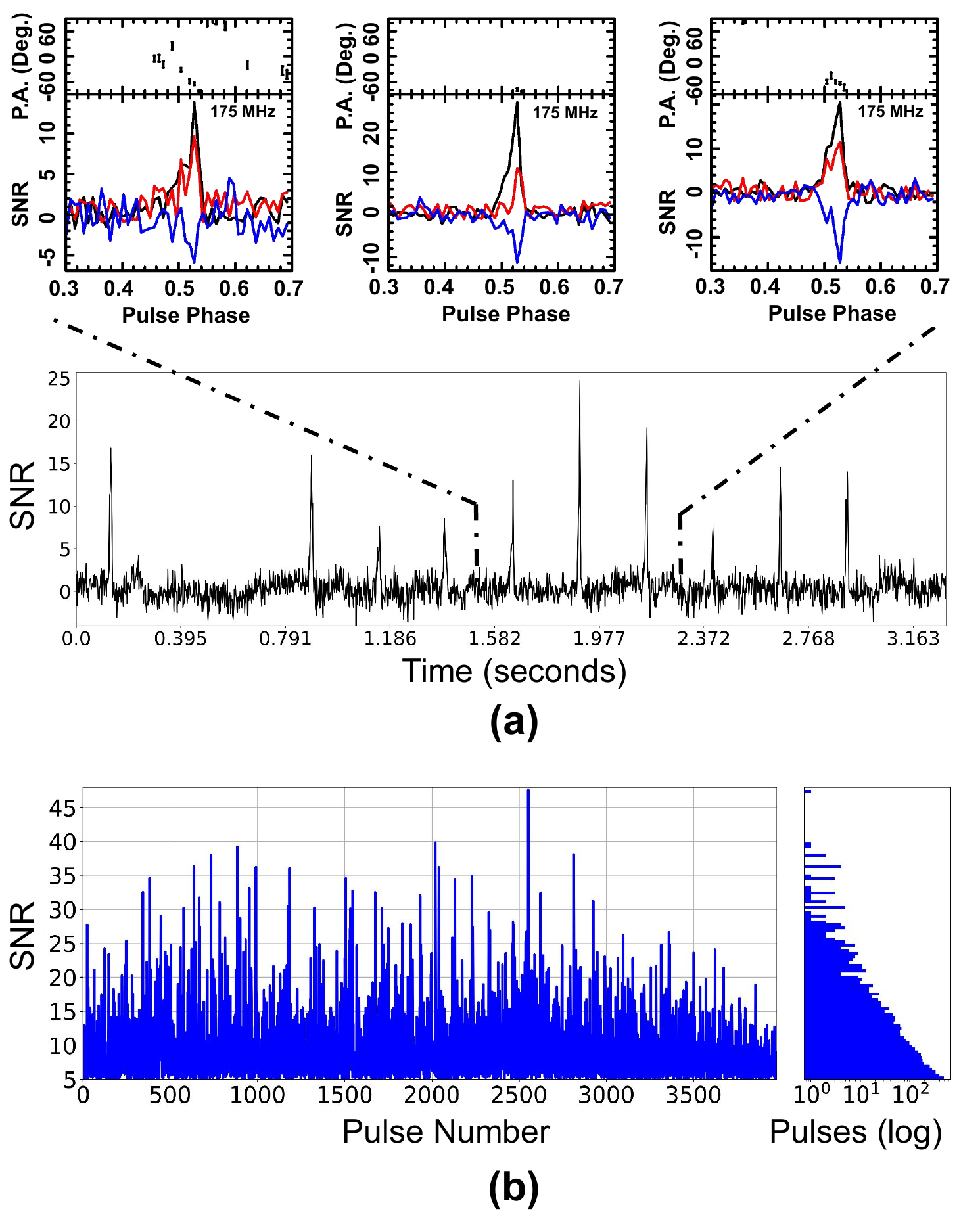} 
    \caption{(a) Three bright single pulses plotted with linear polarisation (Red), Circular polarisation (Blue) [top panel], Single pulse train of 3 second data [bottom panel], (b) 3978 pulses detected above 5 sigma from the J0953+0755 in 40 minutes of observations at 175 MHz on MJD 60994 ) which is routinely observed by the array.  Single pulses were detected at 175 MHz from the observations. } 
    \label{J0953+0755_single_pulses}
\end{figure}

The high-intensity giant pulses from this pulsar were studied for microstructure analysis in the pulse region and to understand the pulse emission mechanism. The morphologies and spectral structure of the pulse were compared with those of FRB and magnetar signals to understand the propagation effects of the source environments \citep{bilous2022dual}. We observed this pulsar on  MJD 60994 with an SNR of 300. \textit{single\_pulse\_search.py} package from the PRESTO was used to conduct a single pulse search on this data and detected 3978 single pulses above an SNR of 5, as shown in Figure \ref{J0953+0755_single_pulses}b. Frequency scrunched time series data of 3.2 seconds with polarisation plot for three consecutive giant single pulses are also demonstrated in \ref{J0953+0755_single_pulses}a. 

\begin{figure}[!ht]
    \centering
    \includegraphics[width=\linewidth]{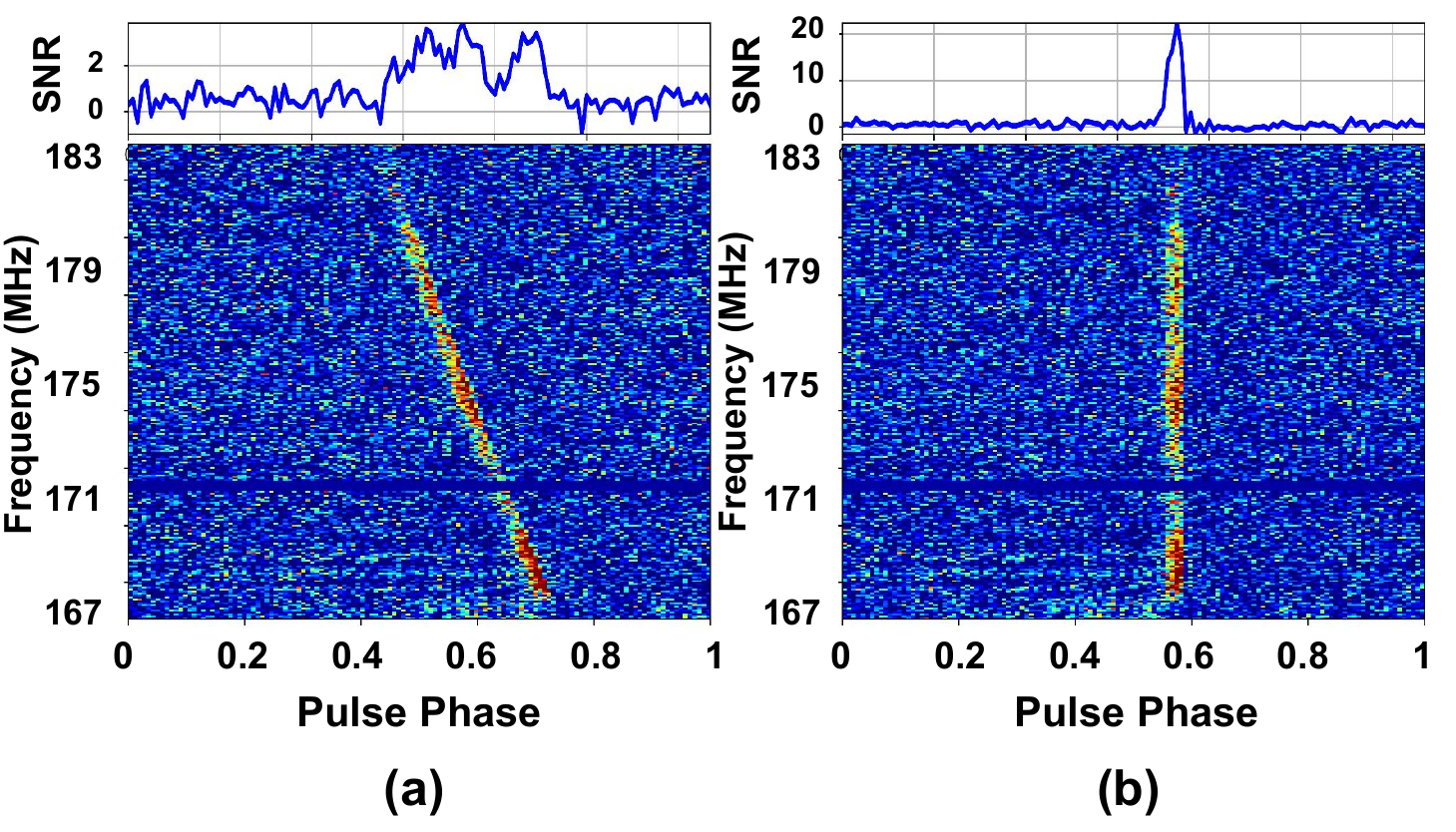}
    \caption{ J0953+0755 Single pulse before and after dispersion correction.}
    \label{J0953+0755_single_pulse_spectrogram}
\end{figure}

Based on the giant pulse event time, the 253-ms length of search mode data was sliced and plotted against frequency with and without de-dispersion, as shown in Figure \ref{J0953+0755_single_pulse_spectrogram}. This work illustrates the features of the processing pipeline for handling and searching radio transients in the observation data. 

\subsection{J0534+2200}
\begin{figure}[ht]
    \centering
    \includegraphics[width=1.0\linewidth]{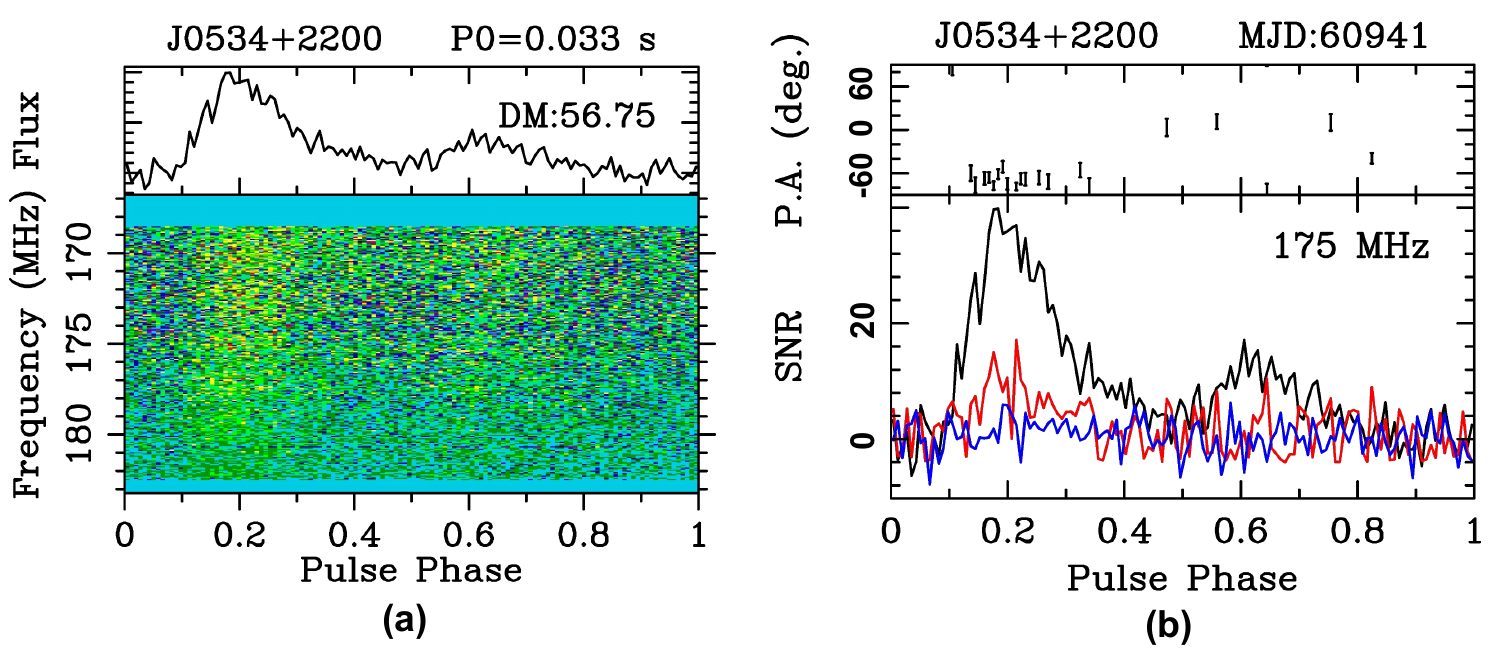}
    \caption{Pulse profile of J0534+2200 (CRAB Pulsar). (a) Pulse intensity across the frequency. (b) Polarisation fraction, total intensity (Black), linear polarisation (Red), circular polarisation (Blue).
    } 
    \label{J0534+2200_profile}
\end{figure}

 The Crab pulsar (PSR B0531+21 or PSR J0534+2200) is a much-studied, young, and energetic pulsar with a period of 33 milliseconds and a dispersion of 56.75 $pc/cm^3$. It is radio-bright, with a spin-down rate  $-3.77535 \times 10^{-10} s^{-2}$  and a flux density of 8 Jy at 150 MHz.  The pulsar is routinely observed to study the giant pulses, nanoshots, and spin period glitches  \citep{singha2022pulsar} it generates. At low frequencies, pulses are scattered, providing a handle to estimate the interstellar medium model \citep{kirsten2019probing}. This pulsar is located at the northern edge of the tile beam, and it has been routinely detected in our observations with SNRs 10 to 16. We have observed this pulsar at 175 MHz and presented the AP profile with intensities across the frequency in Figure \ref{J0534+2200_profile}a. The amount of polarised intensities measured against pulse phase is shown in Figure \ref{J0534+2200_profile}b.

Faraday rotation of this pulsar has been measured using polarisation properties, and an RM of -40.5 $\ pm$ 4.5 $rad/m^2$ is reported \citep{Manchester1971}.  The polarisation variation of this pulsar was used as a probe to study the magnetic field environment of the host nebula over a decade \citep{rankin1988crab}. 

\begin{figure}[ht]
    \centering
    \includegraphics[width=1.0\linewidth]{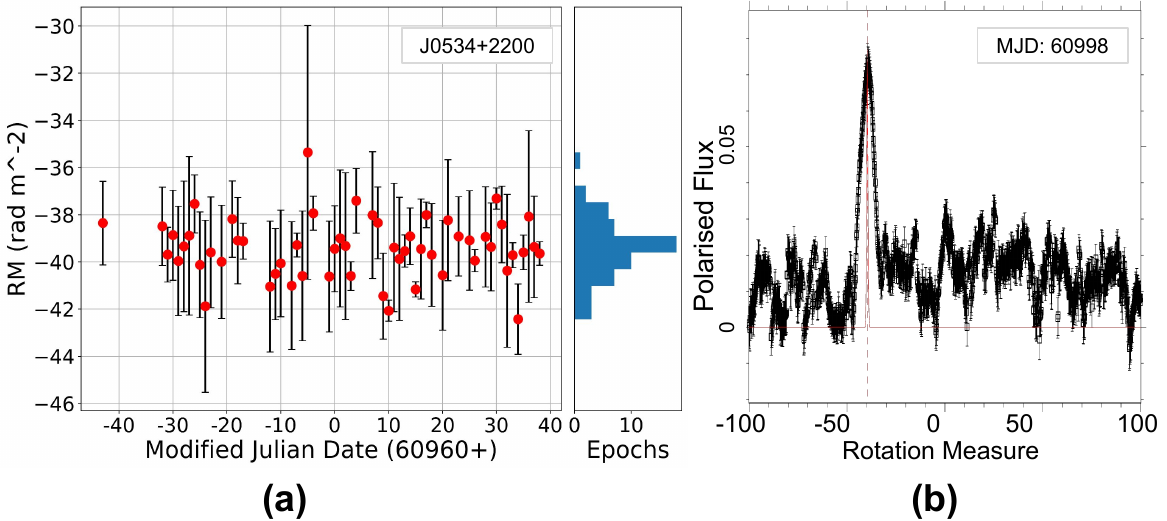}
    \caption{J0534+2200 Rotation measure estimation using rmfit. RM = -39.4 $\mp$ 1.8 $rad/m^2$ } 
    \label{J0534+2200_RM_plot}
\end{figure}

\begin{figure}[ht]
    \centering
    \includegraphics[width=\linewidth]{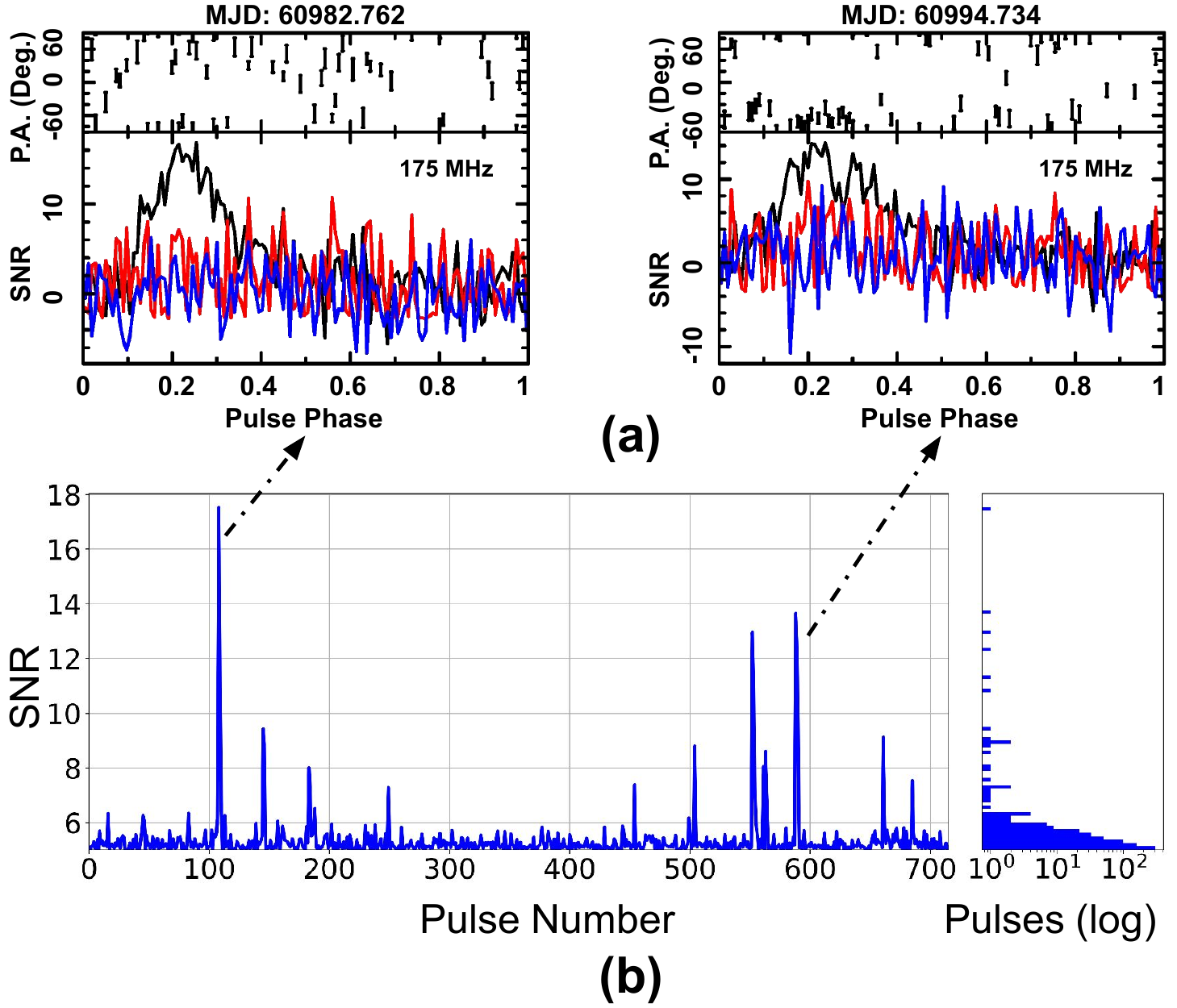}
    \caption{ (a) Two bright single pulses plotted with linear polarisation (Red), Circular polarisation (Blue), (b) 706  Pulses detected above 5 sigma from J0534+2200 in 11.3 hours observations at 175 MHz spanning 18 days) which is routinely observed by the array. J0534+2200 Single pulses detected at 175 MHz from the observations. }
    \label{J0534+2200_single_pulses}
\end{figure}

 Figure \ref{J0534+2200_RM_plot}a presents the Faraday rotation measured for this pulsar over 100 consecutive days in our observation. One of the epoch's \textit{rmfit}  estimate profile is shown in  \ref{J0534+2200_RM_plot}b. The average rotation measure (RM) for the 100 epochs is estimated to be -39.4 $\mp$ 1.8 $rad/m^2$ in this work.

The giant pulses from this pulsar were used for scintillation studies and found to be heavily affected by Kolmogorov turbulence at higher frequencies, which is crucial for placing constraints on the FRB-like searches at higher frequencies \citep{doskoch2024statistical}.
Our observation at 175 MHz over 11.3 hours spanning 18 days in November 2025 detected giant pulses with SNR greater than 5 at a rate of about 1 per minute. Figure \ref{J0534+2200_single_pulses}b shows the counts of detected giant pulses as a function of their SNR. The fractional polarisation measured for two of the giant pulses is shown in Figure \ref{J0534+2200_single_pulses}a.

\begin{figure}[ht]
    \centering
    \includegraphics[width=\linewidth]{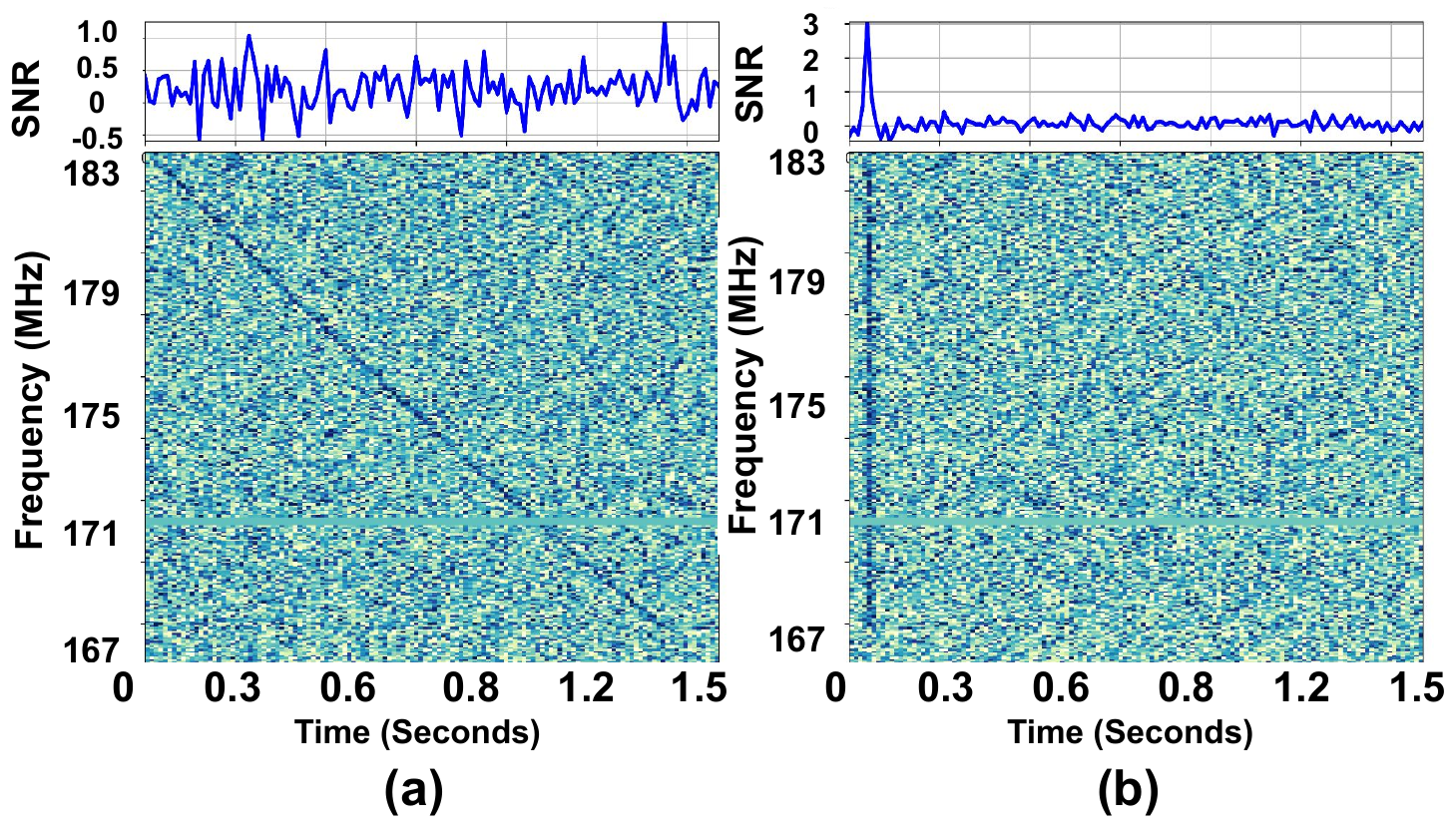}
    \caption{ J0534+2200 Single pulse before and after dispersion correction.}
    \label{J0534+2200_single_pulse_spectrogram}
\end{figure}

\tblu
High-dispersion pulsar signals experience more delay in low-frequency observation. One of the GP of crab pulsar takes almost 1.5 seconds to cross our band of observation, shown in Figure \ref{J0534+2200_single_pulse_spectrogram}a—the expected improvement in SNR after de-dispersion was shown in Figure \ref{J0534+2200_single_pulse_spectrogram}b. The effect of dispersion at low frequencies can be beneficial when searching for high-DM FRB signals.
\tbla
\subsection{J1136+1551}

PSR J1136+1551 (B1133+16) is a normal pulsar with a period of 1.19 seconds and a dispersion measure of 4.841 $pc/cm^3$. This pulsar is an isolated, mode-changing pulsar. It has a complex average profile with two well-separated main components, whose component widths change with frequency, a phenomenon known as radius-to-frequency mapping (RFM) \citep{oswald2019understanding}.

\begin{figure}[ht]
    \centering
    \includegraphics[width=\linewidth]{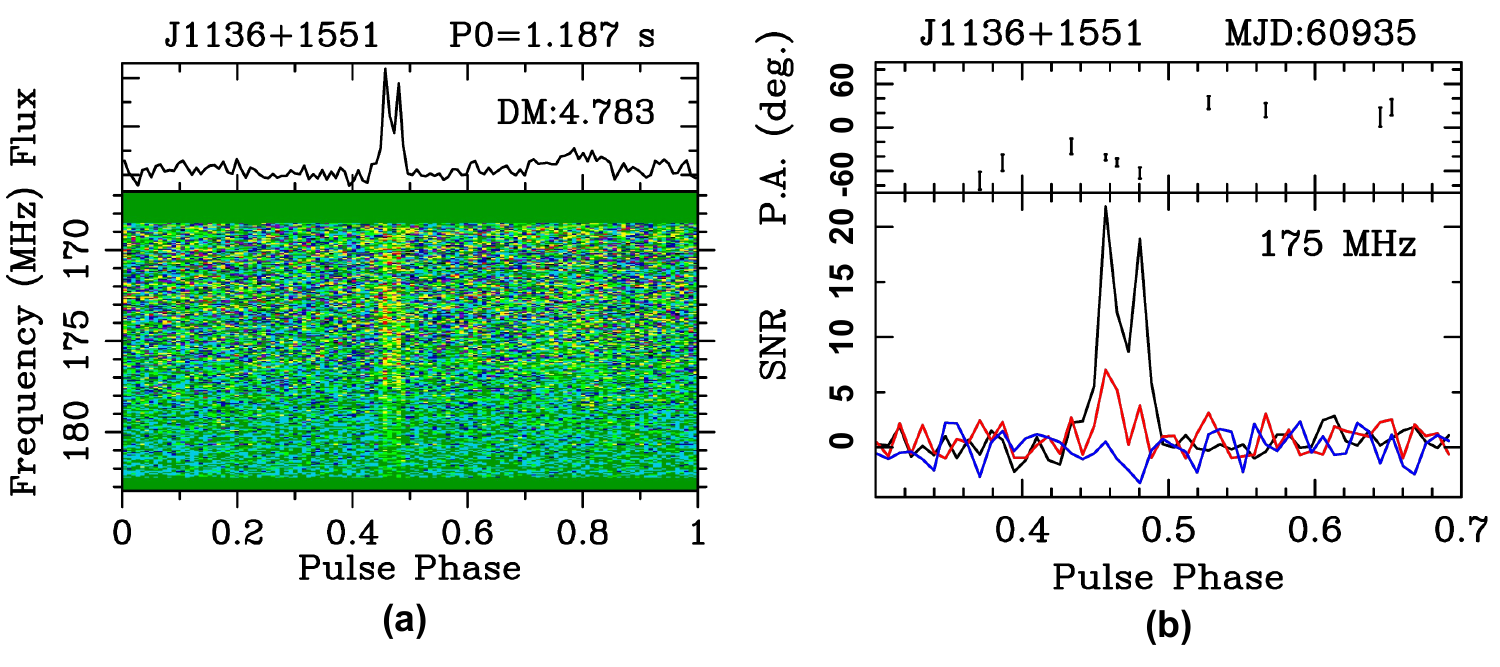}
    \caption{
    Pulse profile of J1136+1551. (a) Pulse intensity across the frequency. (b) Polarisation fraction, total intensity (Black), linear polarisation (Red), circular polarisation (Blue).
    } 
    \label{J1136+1551_profile}
\end{figure}

Figure \ref{J1136+1551_profile}a presents the AP profile with pulse intensities as a function of observation frequency. The corresponding epoch's polarisation fraction of the pulse is shown in Figure \ref{J1136+1551_profile}b.

This pulsar exhibits a sub-pulse drift mode with the bright single pulse emission \citep{tan2024bright} and scintillation \citep{ziwei2022pulsar}. It shows a much larger intensity modulation at low sky frequencies, resulting in narrow, bright emissions that may provide a vital clue to understanding the underlying mechanism of bright-pulse emission \citep{karuppusamy2011low}.

\begin{figure}[ht]
    \centering
    \includegraphics[width=0.9\linewidth]{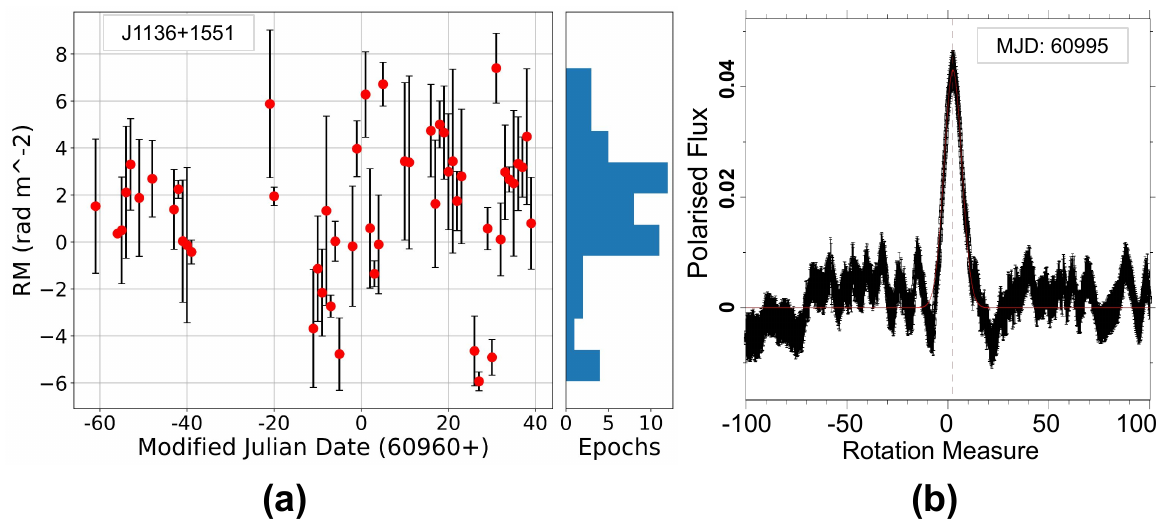}
    \caption{
    J1136+1551 Rotation measure estimation using rmfit. 1.42 $\mp$ 1.9 $rad/m^2$ } 
    \label{J1136+1551_RM_plot}
\end{figure}

 Rotation measure of this pulsar was measured over 100 days of observation, and it was found to vary as shown in Figure \ref{J1136+1551_RM_plot}a. The rmfit profile of one of the high SNR epochs is shown in Figure \ref{J1136+1551_RM_plot}b.

\begin{figure}[ht]
    \centering
    \includegraphics[width=0.9\linewidth]{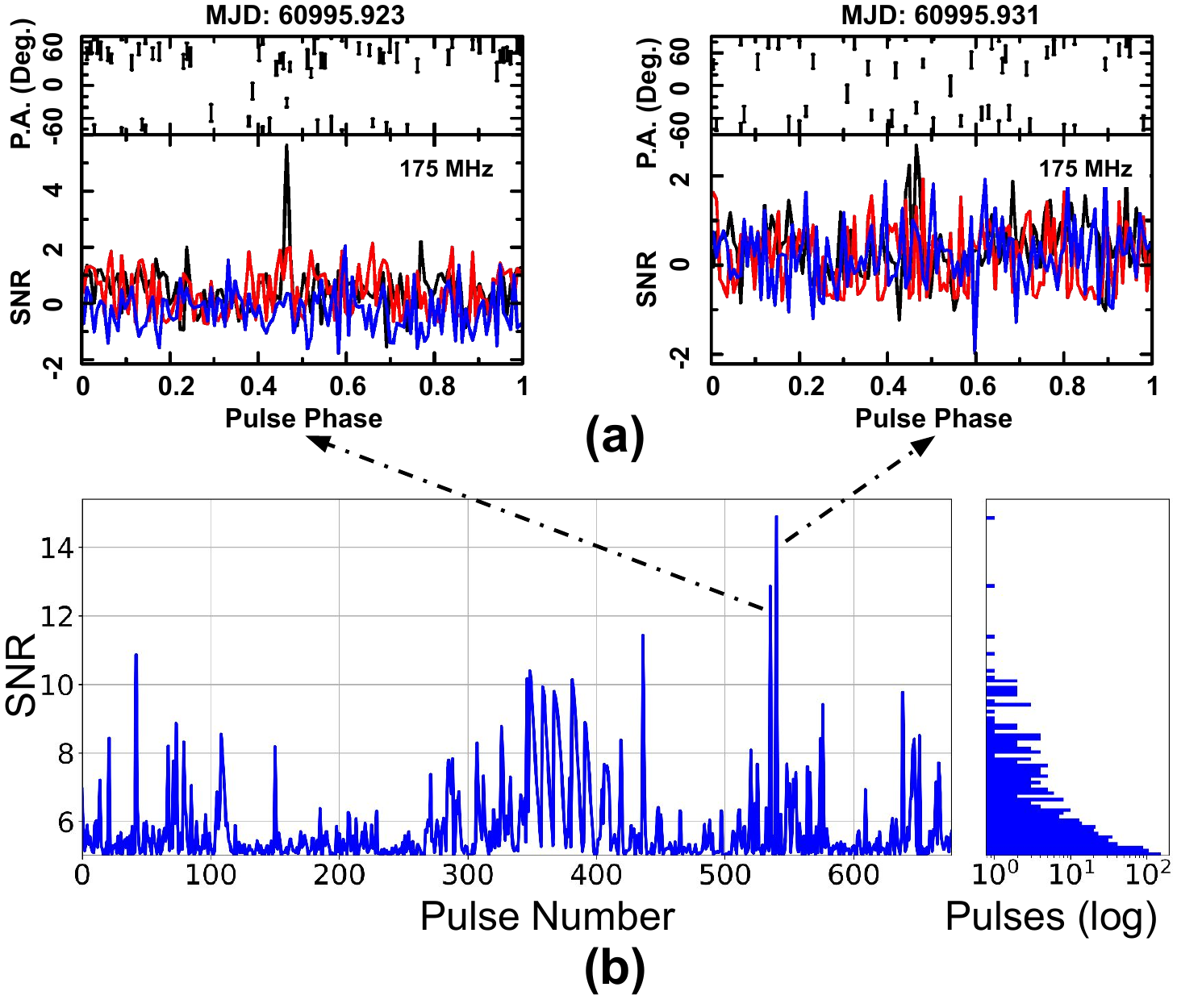}
    \caption{(a) Two bright single pulses plotted with linear polarisation (Red), Circular polarisation (Blue), (b) Single pulse detection of PSR J1136+1551 with S/N more than 5.}
    \label{J1136+1551_single_pulses}
\end{figure}

\tblu
The occurrence of unusually intense single pulses provides a sample of instances in which we can observe frequency evolution along field lines in the magnetosphere, allowing us to develop a description of the shape of the active emission region and build constraints on models of the emission region within the magnetosphere \citep{oswald2019understanding}.\tbla We have conducted a single pulse search in our November 2025 dataset, and found around 690 bright pulses above SNR 5 as shown in Figure \ref{J1136+1551_single_pulses}b.  Two of the bright pulses shown in Figure \ref{J1136+1551_single_pulses}b, and \tblu their total intensity pulse profile was different despite their close occurrence. \tbla %

\subsection{J0837+0610}

PSR J0837+0610 (B0834+06) is an isolated pulsar with a period of 1.27 seconds and a dispersion measure of 12.864 $pc/cm^3$. This pulsar profile has two close pulse components within 10 degrees of pulse longitude, which is also growing towards higher frequencies \citep{johnston2008multifrequency}. 

\begin{figure}[ht]
    \centering
    \includegraphics[width=\linewidth]{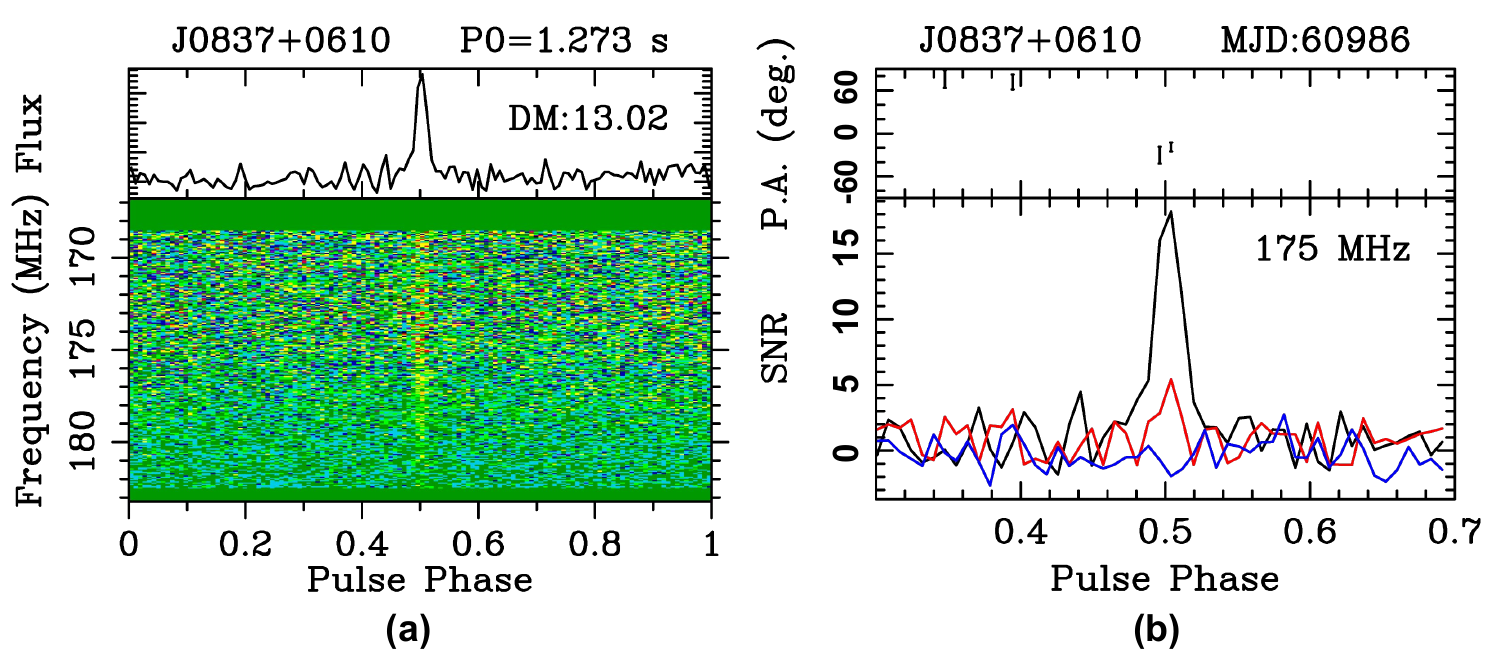}
    \caption{
    Pulse profile of J0837+0610. (a) Pulse intensity across the frequency. (b) Polarisation fraction, total intensity (Black), linear polarisation (Red), circular polarisation (Blue).
    } 
    \label{J0837+0610_profile}
\end{figure}

It exhibits a nulling pulse phenomenon with a fraction of 7-9\% \citep{ritchings1976pulsar, wang2020jiamusi}. The scintillation parameter of this pulsar was studied, and the scattering screen distance was reported in the frequency range 145–155 MHz \citep{ziwei2022pulsar}.

In our observation, this pulsar is detected with an average SNR of 15. Figure \ref{J0837+0610_profile}a, presents the AP profile against observed flux density across the frequency band. The fraction of the polarisation seen in the profile is plotted in Figure \ref{J0837+0610_profile}b.

\subsection{J1921+2153}

PSR J1921+2153 (B1919+21) is a pulsar with a 1.34-second period and a dispersion measure of 12.444 $pc/cm^3$. The rapid change in the fraction of polarisation leads to orthogonal mode jumps in the PA swing and component blending across frequency \citep{johnston2008multifrequency}. The fluctuation spectral analysis of PSR J1921+2153 at 618 MHz revealed a jump in the phase variation across the pulse components, categorised as switching phase-modulated drifting \citep{basu2019classification}. The high time resolution of full polar data was used to study linear polarisation position angle and found that it follows the RVM model \citep{johnston2024thousand}.

\begin{figure}[ht]
    \centering
    \includegraphics[width=\linewidth]{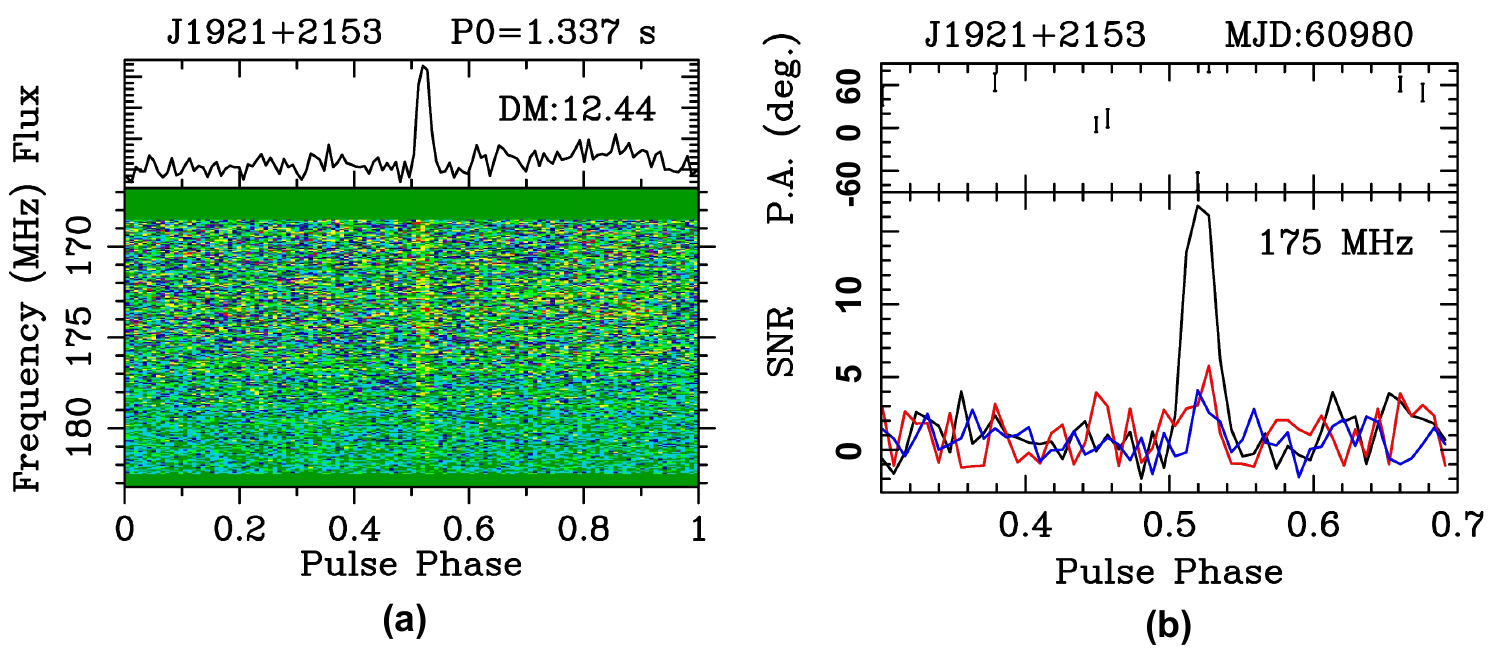}
    \caption{
    Pulse profile of J1921+2153. (a) Pulse intensity across the frequency. (b) Polarisation fraction, total intensity (Black), linear polarisation (Red), circular polarisation (Blue).
    } 
    \label{J1921+2153_profile}
\end{figure}

Most detailed studies of pulsar properties require polarimetric observations. In Figure \ref{J1921+2153_profile}a, we are showing the AP profile from our observation with spectral intensity across time. The corresponding data was plotted with fraction of polarisation in Figure \ref{J1921+2153_profile}b.

\subsection{RFI Mitigation}

Radio frequency interference (RFI) has a significant impact on pulsar observation. Pulsars are weak sources; therefore, RFI is prevalent and reduces the pulse profile's signal-to-noise ratio (SNR). To get a high SNR, we may have to perform RFI cleaning. In our pipeline, we have adopted the RFI removal tool for pulsar archives \textit{Iterative cleaner} \citep{lars2017iterative}, which is based on the surgical cleaner included in the \textit{coast\_guard} \citep{lazarus2020coastguard} pipeline. 

 \subsection{Current status}

\begin{figure}[!h]
    \centering
    \includegraphics[width=\linewidth]{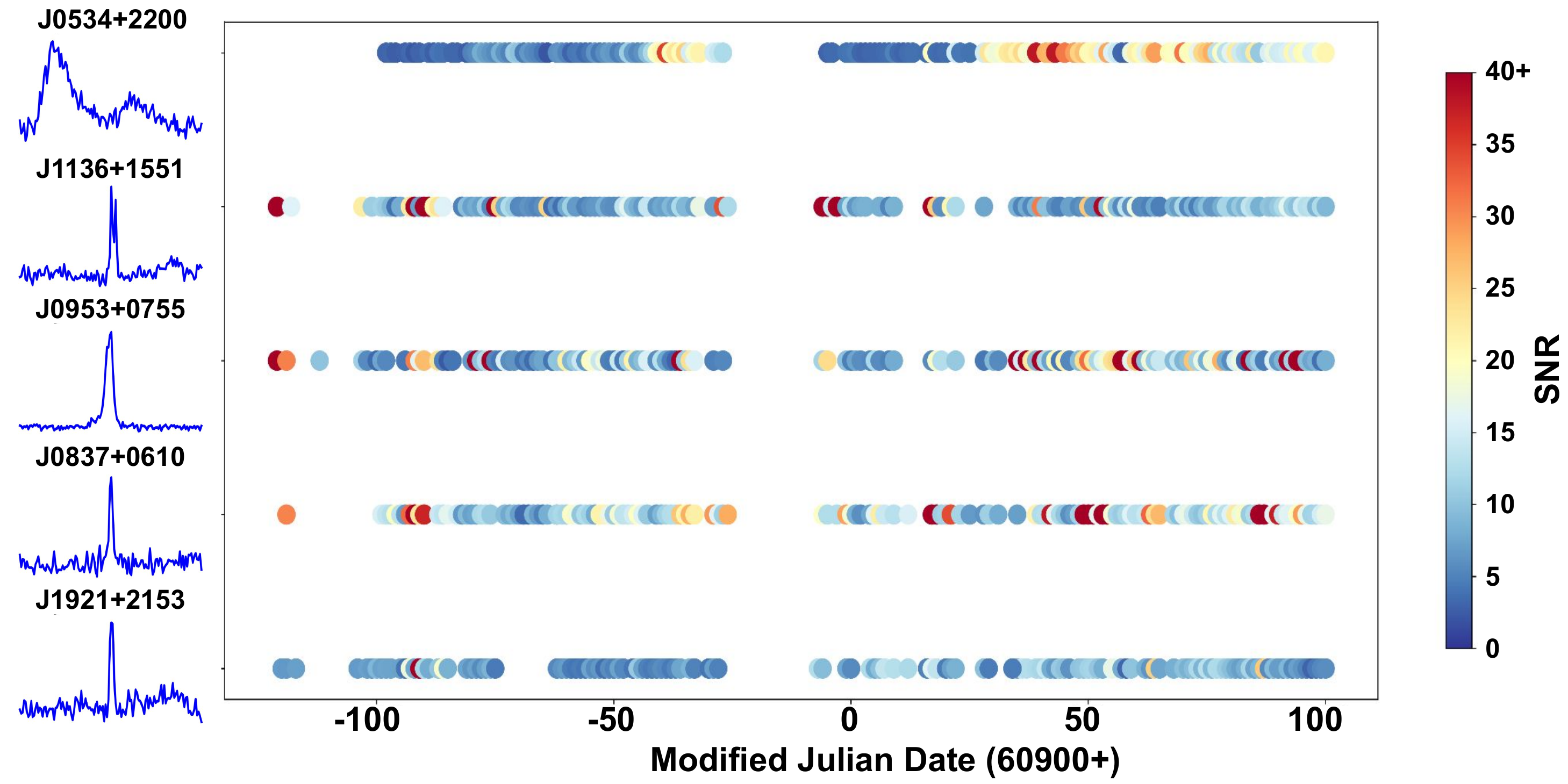}
    \caption{Pulse profiles, observation epochs and the detected SNR for the five bright pulsars routinely observed from MJD 60820 to 61000 by the GBD-DART array.} 
    \label{OBS_cadance}
\end{figure}

For pulsar observations, instrument stability is essential. Two crucial challenges are managing the high data rate and maintaining clock stability over time. The array is routinely monitoring five bright pulsars. Figure \ref{OBS_cadance} shows observation status for five pulsars with observed epoch and corresponding SNRs detected from the analysis. Our observations collect approximately 1.2 TB of data and process it each day on an AMD R9 desktop server. We have a GPS-synchronised clock as a reference system for the backend systems.

\begin{figure}[ht]
    \centering
    \includegraphics[width=\linewidth]{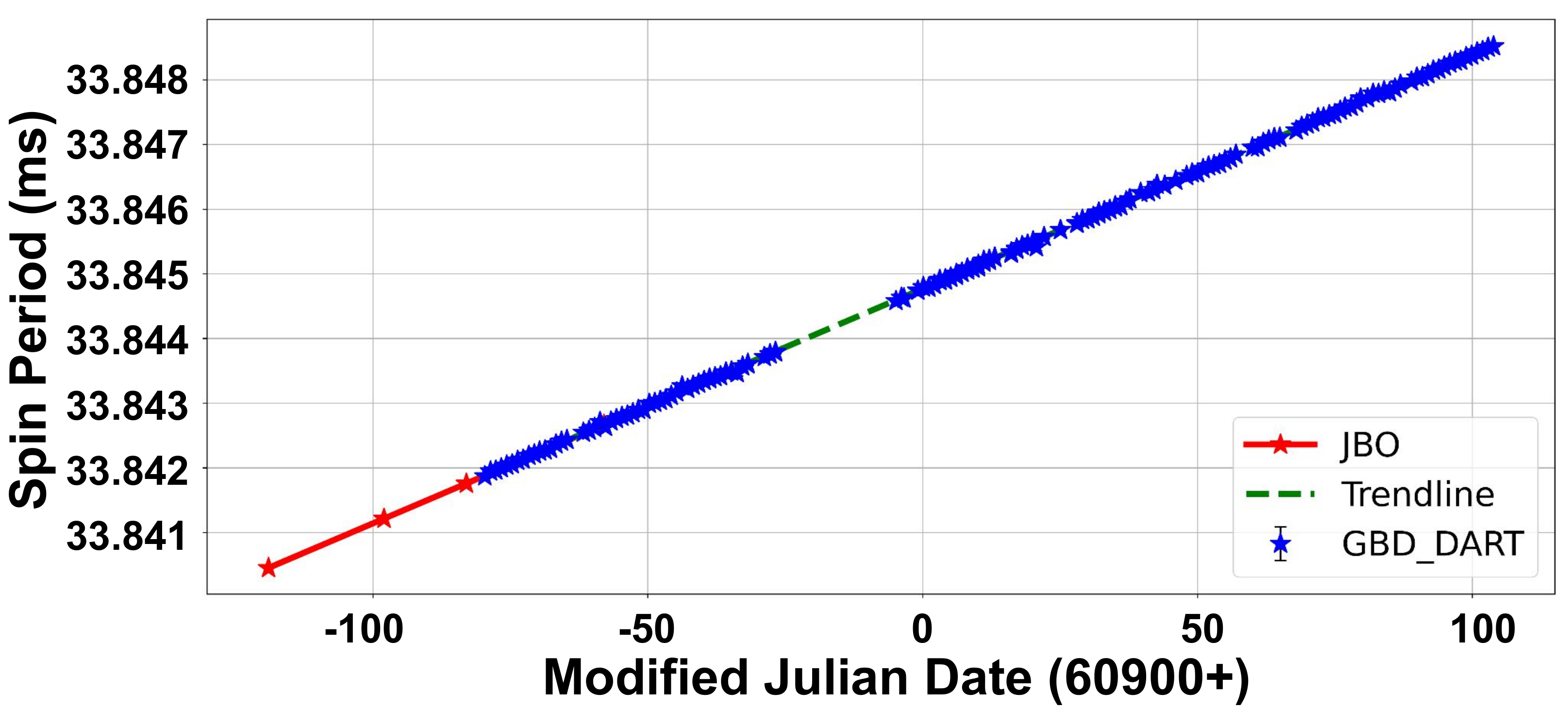} 
    \caption{Spin period of PSR J0534+2200 estimation from our observation between MJD 60820 and 61000 (Blue). Monthly ephemerides data for this pulsar from JBO (Red) and extended trend line from the data (green).   }  
    \label{J0534+2200_P0_P1}
\end{figure}

The Crab pulsar is one of the regularly observed sources in the array; we report here its spin-down obtained over nearly 200 days of observation.   The estimated spin period of PSR J0534+2200, as determined using \textit{PDMP} over approximately 200 days, is plotted against MJD in Figure \ref{J0534+2200_P0_P1}. We utilised the Jodrell Bank Observatory (JBO) CRAB pulsar monthly ephemeris data \citep{lyne199323} to verify the spin-down trend line of the spins, which closely match at the first order with our estimates.

\section {conclusion}
We have developed a new pulsar data processing pipeline for the GBD-DART array. The pipeline consists of both hardware and software modules. In-house-built analog and digital systems are used in the hardware pipeline, and custom-developed Python scripts and standard pulsar tools are employed to implement the various stages of the software pipeline. The entire pipeline was validated with the emulated and simulated pulse signals. The online monitoring feature is incorporated, which provides course-averaged data to monitor the array's health. A transient buffer is implemented to transfer raw data upon receiving a trigger for the scheduled pulsar observation and to respond to transient events, such as the FRB detection alert from the Virtual Observatory Event (VOEvent). The pipeline offers coherent and incoherent mode dedispersion. Standard open source RFI mitigation tools were used in this pipeline. We employ an incoherent mode to assess the quantity of good data mitigating RFI (rfifind) in the search-mode fits data. Search-mode data are routinely backed up to facilitate the search for radio transients and single pulses from pulsars. We have presented results from nearly 200 days of observing five bright pulsars: J0953+0755, J0534+2200, J1136+1551, J0837+0610, and J1921+2153. The results include average pulse profiles with polarisation for all the pulsars, results from an RM estimate and a single-pulse study for J0953+0755, J0534+2200, and J1136+1551, and spin-down monitoring results for the Crab pulsar J0534+2200.

Additionally, we have provided the cadence details and SNR for all five pulsars. For archives, we choose to use data formats widely used in the pulsar community. When there is a particular interest, the raw ADC data can also be retained in the DADA format. The data products are stored in folded archives in search mode, in a PSRFITS format compatible with PSRCHIVE. Standard timing tools, such as the tempo and tempo2, can also be used to process the essential information required for pulsar timing. In our observations, CPU multi-threading reduced data processing time nearly in proportion to the observation time. To obtain the intrinsic polarisation characteristics of the pulsar, we have chosen to focus on polarisation calibration in the future. Beam steering, either in analog or digital mode, and broadband observation are planned for the future.

\subsection*{Disclosures}
The authors declare there are no financial interests, commercial affiliations, or other potential conflicts of interest that have influenced the objectivity of this research or the writing of this paper.

\subsection*{Conflicts of Interest}
None.

\subsection*{Code, Data, and Materials Availability}


The data, materials and software code presented in this article are publicly available in \url{https://github.com/Arul16psp05/GBD_DART_lite.git} at  \url{https://doi.org/10.5281/zenodo.18184048}

\subsection*{Acknowledgments}
We acknowledge the multiple technical consultations with Indrajit Vittal Barve and Shaik Sayuf. We acknowledge the involvement of our colleagues, staffs and security at the Gauribidanur Observatory for their support during the observation, particularly Narendra for the power maintenance. Additionally, we thank the RRI electrical team support on UPS system. We are also grateful to our EEG colleagues for their various forms of support and valuable conversations that aided this work. We thank the Raman Research Institute for supporting this development work. We also thank the Christ University for recognising the research aspect of this effort. Also, we acknowledge the use of RRI library provided access to the Grammarly \citep{fitria2021grammarly} tool and publicly available QuillBot \citep{fitria2021quillbot} to correct grammar in the manuscript.

\section*{Appendix}

\begin{figure}[!h]
    \centering
    \includegraphics[width=\linewidth]{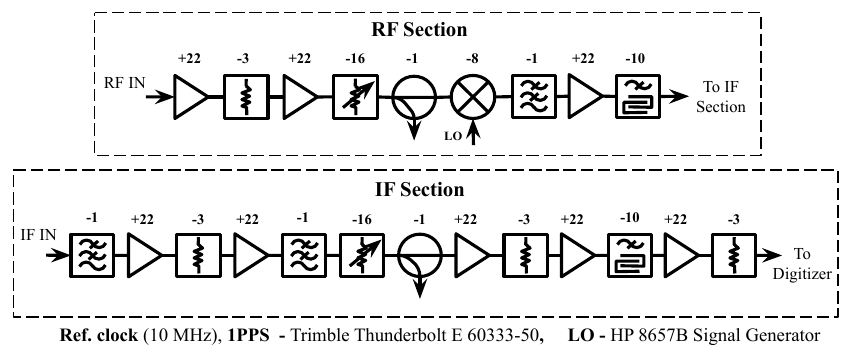}
    \caption{Signal flow of the PDR RF and IF sections } 
    \label{PDR_RF_IF}
\end{figure}

\end{document}